\documentclass[12pt]{article}
\usepackage{epsfig}
\usepackage{graphicx}
\usepackage{amssymb}
\usepackage{amsmath}
\usepackage{amsfonts}
\usepackage[dvips]{color}

\newcommand{\R}{\mathbb{R}}
\newcommand{\C}{\mathbb{C}}

\newcommand{\fg}{\mathfrak{g}}

\newcommand{\fS}{\mathfrak{S}}

\newcommand{\cC}{\mathcal{C}}

\newcommand{\cH}{\mathcal{H}}

\newcommand{\cP}{\mathcal{P}}

\newcommand{\cT}{\mathcal{T}}

\newcommand{\be}{\begin{equation}}
\newcommand{\ee}{\end{equation}}
\newcommand{\bea}{\begin{eqnarray}}
\newcommand{\eea}{\end{eqnarray}}
\newcommand{\nn}{\nonumber}
\newcommand{\kt}{\rangle}
\newcommand{\br}{\langle}

\newcommand{\ed}{\end{document}}

\newcommand{\bi}{\begin{itemize}}
\newcommand{\ei}{\end{itemize}}

\newcommand{\etap}{{\eta_{_+}}}

\newcommand{\bce}{\begin{center}}
\newcommand{\ece}{\end{center}}

\begin{document}

\title{Non-Hermitian Hamiltonians with a Real Spectrum
and Their Physical Applications}

\author{Ali~Mostafazadeh\thanks{E-mail address:
amostafazadeh@ku.edu.tr}
\\
Department of Mathematics, Ko\c{c} University,\\ 34450 Sariyer,
Istanbul, Turkey}

\date{ }
\maketitle

\begin{abstract}

We present an evaluation of some recent attempts at understanding
the role of pseudo-Hermitian and $\cP\cT$-symmetric Hamiltonians in
modeling unitary quantum systems and elaborate on a particular
physical phenomenon whose discovery originated in the study of
complex scattering potentials.\vspace{2mm}

\vspace{2mm}

\end{abstract}
\vspace{5mm}

%\np

\section{Introduction}
The use of non-Hermitian operators in theoretical physics has a long
history \cite{review}. These operators are traditionally employed in
the effective description of physical systems displaying decay or
dissipative behavior. The main quality of non-Hermitian operators
that motivated these applications is that a generic non-Hermitian
operator has complex eigenvalues whose imaginary part may be
associated with decay rates. This property is not however common to
all non-Hermitian operators; there is a class of non-Hermitian
operators that similarly to Hermitian operators have a real
spectrum. During the past ten years or so, these operators have been
the focus of an intensive research activity particularly following
the work of Bender and Boettcher \cite{bender-prl} on
$\cP\cT$-symmetric Hamiltonian operators. The possibility that these
operators can have a purely real spectrum has led to the conjecture
that one can actually use them to describe unitary quantum systems.
In the present article, I elaborate on the physical significance of
the above conjecture and draw the attention of the reader to a
physical phenomenon that has been recently discovered in an attempt
to understand a class of non-Hermitian Hamiltonian operators with a
continuous spectrum. To the best of my knowledge, this is the first
example of a physical effect whose discovery has its origin in the
study of non-Hermitian Hamiltonians possessing a real spectrum.

\section{Basic Ingredients of the Formalism}

For a system having the real line $\R$ as the configuration space,
the parity and time reversal operators are respectively defined by
$\cP \psi(x):=\psi(-x)$ and $\cT\psi(x):=\psi(x)^*$, where $\psi$ is
an arbitrary complex-valued function typically belonging to
    $L^2(\R):=\left\{\psi:\R\to\C~\left|~\int_{-\infty}^\infty
   |\psi(x)|^2\: dx<\infty\right.\right\}.$
A linear operator $H$ acting on $L^2(\R)$ is said to posses
\textbf{$\cP\cT$-symmetry} if $[H,\cP\cT]=0$. If there is a complete
set of eigenvectors $\psi_n$ of $H$ that are $\cP\cT$-invariant,
i.e., $\cP\cT\psi_n=\psi_n$, then $H$ is said to have an
\textbf{exact $\cP\cT$-symmetry}. The latter is a very strong and
difficult-to-check condition on $H$.

Because $\cP\cT$ is an antilinear operator, $\cP\cT$-symmetry
implies that the spectrum of $H$ is symmetric about the real axis.
In particular, complex eigenvalues come in complex-conjugate pairs.
This is actually very easy to show. Similarly, it is easy to show
that exact $\cP\cT$-symmetry implies the reality of all the
eigenvalues. What is by no means easy to show is whether a given
operator possesses exact $\cP\cT$-symmetry. To do this one must
first establish the existence of a  $\cP\cT$-invariant set of
eigenvectors of $H$ and prove their completeness.

In order to introduce a notion of completeness, one must adopt a
particular notion of convergence on the function space where $H$
acts. If this is $L^2(\R)$, one usually takes the standard
$L^2$-inner product: $\br\psi|\phi\kt:=\int_{-\infty}^\infty
\psi(x)^*\phi(x)\,dx$ to define a norm, namely
$\parallel\psi\parallel:=\sqrt{\br\psi|\psi\kt}$, and use the latter
to determine the convergence of sequences. We shall denote by $\cH$
the Hilbert space obtained by endowing $L^2(\R)$ with this inner
product.

If $\{\xi_n\}$ is a sequence in $\cH$, and for every $\psi\in\cH$
there are complex numbers $c_n$ such that the series
$\sum_{n=1}^\infty c_n\xi_n$ converges to $\psi$, i.e.,
$\lim_{N\to\infty}\parallel\psi-\sum_{n=1}^Nc_n\psi_n\parallel=0$,
we say that $\{\xi_n\}$ is a basis of $\cH$. It turns out that the
notion of ``completeness'' is a stronger condition on a sequence
$\{\xi_n\}$. It is equivalent to the existence of a complementary
sequence $\{\zeta_n\}$ in $\cH$ that satisfies
$\br\zeta_m|\xi_n\kt=\delta_{mn}$. $\{\zeta_n\}$ is then also a
basis and $\{(\xi_n,\zeta_n)\}$ is called a biorthonormal system,
\cite{review}. The bases $\{\xi_n\}$ that have this property are
called \textbf{Riesz bases}. Every orthonormal basis
$\{\varepsilon_n\}$ is clearly a Riesz basis, because
$\{(\varepsilon_n,\varepsilon_n)\}$ is a biorthonormal system. We
say that a linear operator $H:\cH\to\cH$ is \textbf{diagonalizable}
if $H$ has a set of eigenvectors $\psi_n$ that form a Riesz basis,
i.e., it has a complete set of eigenvectors. In this case the
complementary (biorthonormal) basis $\{\phi_n\}$ associated with
$\{\psi_n\}$ are eigenvectors of the adjoint of $H$ that we denote
by $H^\dagger$. A precise definition of $H^\dagger$ is given in
\cite{review}. Here we suffice to mention that for any $\psi$ and
$\phi$ taken respectively from the domains of $H$ and $H^\dagger$,
we have $\br\psi|H\phi\kt=\br H^\dagger\psi|\phi\kt$. We will also
refer to $\{(\psi_n,\phi_n)\}$ as a biorthonormal eigensystem for
$H$.

An operator $H:\cH\to\cH$ is said to be \textbf{Hermitian} or
self-adjoint if $H^\dagger=H$. This means that for every $\psi,\phi$
in the domain of $H$, $\br\psi|H\phi\kt=\br H\psi|\phi\kt$. Much of
the confusion in the study of $\cP\cT$-symmetric Hamiltonians may be
traced to the misconception that the notion of ``Hermiticity'' can
be defined independently of the choice of the inner product of
$\cH$. Many authors follow the naive and unjustified practice of
choosing a preferred basis (such as the position basis) in $\cH$,
represent the operator $H$ using a matrix $\underline{H}$ in this
basis, and define the Hermiticity condition as the requirement that
the transpose of $\underline{H}$ be equal to its complex-conjugate,
$\underline{H}^t=\underline{H}^*$. This is OK only if the basis one
works with is an orthonormal basis. But to determine the
orthonormality of a basis one needs to use the inner product of
$\cH$. \emph{The term ``Hermitian operator'' is meaningless unless
one specifies this inner product}. Clearly, different choices for
the inner product lead to different notions of ``Hermiticity.''

The mystery underlying the reality of the spectrum of the
$\cP\cT$-symmetric Hamiltonian operators such as $p^2+ix^3$ is
unraveled once one recognizes that these operators are actually
Hermitian with respect to a nonstandard inner product \cite{p2}. It
is misleading to claim that Hermiticity is an unphysical condition,
and hence it must be replaced by the physical condition of
$\cP\cT$-symmetry which represents space-time reflection symmetry
\cite{bbj}. It is in fact easy to show that one can never avoid the
requirement of the Hermiticity of observables, because this is a
necessary condition for the reality of expectation values
\cite{review}. \emph{What the recent developments have revealed is
the possibility of employing nonstandard inner products in quantum
mechanics}. This summarizes the main conceptual outcome of more than
ten years of intensive research on this subject.

This result has a number of important implications \cite{review}.
    \begin{enumerate}
    \item Contrary to initial expectations, $\cP\cT$-symmetry does not play
    any distinctively important role. Any $\cP\cT$-symmetric or
    non-$\cP\cT$-symmetric operator that has a real spectrum and a complete
    set of eigenvectors can serve the same purpose.
    These operators (with proper extension of the notion of completeness to
    the cases that the spectrum possesses a continuous part) can be related
    to Hermitian operators by a similarity transformation \cite{p2}. Hence
    they are quasi-Hermitian \cite{quasi}. Furthermore, every quasi-Hermitian
    operator $H$ has an exact symmetry generated by an antilinear
    operator $\fS$ that is an involution, i.e., $H$ has a
    $\fS$-invariant complete set of eigenvectors and $\fS^2=1$,
    \cite{p3,jmp-2003}. Clearly, $\cP\cT$ is just a particular example
    of $\fS$. Other examples of antilinear symmetries have been
    considered in \cite{jpa-2008a}.

    \item In contrast to the initial expectations \cite{bbj}, the use of
    $\cP\cT$-symmetric Hamiltonians does not lead to a genuine extension
    of quantum mechanics. Rather, it provides a new representation of the
    same theory where the physical Hilbert space is
    defined using the new inner product. The latter can be most
    straightforwardly constructed as follows. First, one recalls that every
    inner product has the form $\br\psi,\phi\kt_{\etap}:=\br\psi|\etap\phi\kt$
    for some positive automorphism (a bounded invertible linear
    operator) $\etap$ called a \textbf{metric operator}. The inner
    products $\br\cdot,\cdot\kt_{\etap}$ with respect to which
    a quasi-Hermitian operator $H$ is Hermitian are given by
    metric operators satisfying the pseudo-Hermiticity
    condition \cite{p1}:
        \be
        H^\dagger=\etap H\etap^{-1}.
        \label{ph}
        \ee
    The physical Hilbert space is then defined using the inner product
    $\br\psi,\phi\kt_{\etap}$, \cite{jpa-2003,jpa-2004b}. We will
    denote the resulting space by $\cH_\etap$.

    \item To determine the physical meaning of a given
    quasi-Hermitian operator $H$, one needs to choose an admissible
    inner product (that renders $H$ Hermitian) and map the latter to an
    equivalent Hermitian operator. A canonical choice is
    $h_\etap:=\etap^{\!\!\!\!1/2}H\,\etap^{\!\!\!\!-1/2}$. One can show
    that as a linear operator mapping $\cH_\etap$ to $\cH$ the
    operator $\etap^{\!\!\!\!1/2}$ is a unitary operator \cite{jpa-2003}.
    Therefore the Hilbert space-Hamiltonian pairs $(\cH_{\etap},H)$
    and $(\cH,h_\etap)$ are unitary-equivalent; they describe the same
    physical system \cite{cjp-2004b}. We will refer to them as the
    \textbf{pseudo-Hermitian} and \textbf{Hermitian representations} of the
    system, respectively.

    \item Some of the notions developed in the study of
    $\cP\cT$-symmetric Hamiltonians do not actually play a
    fundamental role. The primary example is the $\cC$ operator that
    is used as a tool for specifying a particular example
    of the inner products $\br\cdot,\cdot\kt_\etap$ called the
    $\cC\cP\cT$-inner product, \cite{bbj}. As shown in
    \cite{jmp-2003,review}, this inner product corresponds to the choice
    $\etap=\cP\cC$, where $\cC$ is required to fulfil \cite{bbj}
        \be
        \cC^2=1,~~~~[\cC,H]=0,~~~~[\cC,\cP\cT]=0.
        \label{C}
        \ee
    According to the prescription given in
    \cite{bbj,bender-review},
    one must first solve the operator equations (\ref{C}) and
        \be
        \cC=e^Q\cP,
        \label{ansatz}
        \ee
    for a Hermitian operator $Q$, substitute the result in (\ref{ansatz})
    to determine $\cC$, and then construct the $\cC\cP\cT$-inner product
    which actually coincides with $\br\cdot,\cdot\kt_{e^{-Q}}$. Therefore,
    this procedure provides means for computing a metric operator of the form
    $\etap=e^{-Q}$.\footnote{Because metric operators are by definition
    positive, they can always be expressed in this form.} In fact,
    all the quantities of interest, for example the equivalent Hermitian
    Hamiltonian, physical obsevables, and expectation values, only involve
    the metric operator. Therefore, the $\cC$ operator has a secondary role
    as far as the physical aspects are concerned.

    An alternative procedure to the one based on the $\cC$ operator,
    that actually gives the most general admissible inner product, is
    to solve a single operator equation, namely (\ref{ph}), for $\etap$.
    Different methods of solving this equation are
    discussed in \cite{review}. In particular, there is a highly
    effective method of dealing with this equation that involves
    expressing it as a differential equation for the kernel $\br
    x|\etap|y\kt$, \cite{jmp-2006a}. The approach based on the
    pseudo-Hermiticity relation (\ref{ph}) avoids dealing with a
    $\cC$ operator and its defining equations (\ref{C}) and
    (\ref{ansatz}), \cite{review}. Therefore, it is more direct.
    \end{enumerate}

\section{From Formalism to Applications}

Pseudo-Hermitian representation of quantum mechanics and the
techniques developed in the course of its investigation have found
applications in a variety of subjects \cite{review}. Here we wish to
discuss a rare example of a physical phenomenon whose discovery
originated in trying to address the problem of the existence of a
metric operator for a class of non-Hermitian operators with a real
and continuous spectrum \cite{jpa-2009}.

As we explained in Section~2, a linear operator $H$ that satisfies
the pseudo-Hermiticity relation (\ref{ph}) acts as a Hermitian
operator in the Hilbert space $\cH_\etap$. This implies that $H$ has
a real spectrum and a complete set of eigenvectors. Clearly, these
two properties are independent; there are operators with a complete
set of eigenvectors (diagonalizable operators) that lack a real
spectrum, and there are operators with a real spectrum that are not
diagonalizable.

Usually the lack of diagonalizability is associated with the
presence of exceptional points. These are degeneracies at which both
the eigenvalues and eigenvectors coalesce. Exceptional points have
interesting physical implications \cite{eps}. They may appear for
operators acting in finite or infinite-dimensional Hilbert spaces.
There is also another less-known obstruction to the
diagonalizability of non-Hermitian operators called spectral
singularities. These may occur for non-Hermitian operators whose
spectrum has a continuous part (Hence the space in which they act is
necessarily infinite-dimensional.)

Spectral singularities were discovered in the mid 1950's by Naimark
\cite{naimark} and studied thoroughly by mathematicians
\cite{ss-math}. In the context of recent study of $\cP\cT$-symmetric
Hamiltonians, the possibility of the presence of a spectral
singularity was initially noted by Samsonov \cite{samsonov} who
following the work of Naimark \cite{naimark,naimark-book} only
considered models defined on the half-line. In
Ref.~\cite{jpa-2006b}, I have worked out the computation of a metric
operator and the corresponding Hermitian Hamiltonian for
$H=-\frac{d^2}{dx^2}+z\delta(x)$ where $z$ is a complex coupling
constant, $x$ takes values in the whole real line, and $\delta(x)$
is the Dirac delta function. For this model a spectral singularity
manifests itself as an obstruction for the construction of a
biorthonormal eigensystem for the case that $z$ is purely imaginary.
In Ref.~\cite{jpa-2009} we explore the mechanism by which spectral
singularities spoil the completeness of the eigenfunctions for a
general complex scattering potential. We also offer a detailed
investigation of the spectral singularities for the Hamiltonians of
the form $H=-\frac{d^2}{dx^2}+z_-\delta(x+a)+z_+\delta(x-a)$, with
$a\in\R^+$ and $z_\pm\in\C$, which include as a special case the
$\cP\cT$-symmetric Hamiltonians corresponding to the choice
$z_-=z_+^*$, \cite{demiralp2}.

Ref.~\cite{p89} provides a physical interpretation for spectral
singularities. It turns out that spectral singularities correspond
to the energies where both the left and right transmission and
reflection coefficients diverge. In other words, they are associated
with resonances having a zero width. This resonance phenomenon may
be realized in an electromagnetic waveguide modeled using the
$\cP\cT$-symmetric barrier potential:
    \be
    H=-\frac{d^2}{dx^2}+v_{a,\zeta}(x),~~~~
    v_{a,z}(x):=\left\{\begin{array}{ccc}
    i\zeta &{\rm for}& -a<x<0,\\
    -i\zeta &{\rm for}& 0<x<a,\\
    0 &{\rm for}& x=0~{\rm or}~|x|>a,\end{array}\right.~~~~
    \zeta\in\R,~~~~a\in\R^+.
    \nn
    \ee
It implies that at the energies (frequencies) of spectral
singularities, such a waveguide may be used as a resonator. This is
a new physical effect that awaits an experimental confirmation.

In the next section I explore the spectral singularities of an
imaginary delta-function potential. This is one of the simplest
exactly solvable complex potentials that one can consider. Yet the
possibility that this potential might involve spectral singularities
was noted quite recently \cite{jpa-2006b}. This is mainly because
spectral singularities do not play an important role unless one
attempts at constructing a metric operator for the system. The
latter could be realized only after the development of a particular
method of computing metric operators \cite{p2,jpa-2003,jpa-2004b}
that is called ``Spectral Method'' in \cite[\S 4]{review}.

\section{Spectral Singularities}

Consider a complex potential $v:\R\to\C$ such that
$\int_{-\infty}^\infty(1+|x|)|v(x)|dx<\infty$. Then the spectrum of
the Schr\"odinger operator $H=-\frac{d^2}{dx^2}+v(x)$ that is
defined over the whole real line has a continuous part. Suppose for
simplicity that the spectrum is just $[0,\infty)$. The (generalized
or scattering) eigenvalues $E$ are doubly degenerate and the
corresponding eigenfunctions have the following asymptotic behavior.
    \be
    \psi_k^{\fg}(x)\to A_\pm^\fg e^{ikx}+B_\pm^\fg
    e^{-ikx},~~~~{\rm for}~~~~x\to\pm\infty.
    \label{asmp}
    \ee
Here $\fg$ is a degeneracy label taking values 1 and 2,
$k:=\sqrt{E}$, and $A_\pm^\fg$, $B_\pm^\fg$ are possibly
$k$-dependent complex coefficients. One can use the eigenvalue
equation for $H$ to relate $A_+^\fg$ and $B_+^\fg$ to $A_-^\fg$ and
$B_-^\fg$. The result can be expressed in terms of a transfer matrix
$\mathbf{M}$ that by definition fulfils
    \be
    \left(\begin{array}{c} A_+^\fg\\B_+^\fg\end{array}\right)=\mathbf{M}
    \left(\begin{array}{c} A_-^\fg\\B_-^\fg\end{array}\right).
    \label{transfer}
    \ee
It is easy to show that $\det \mathbf{M}=1$, \cite{jpa-2009}.

A particularly useful choice for a pair of eigenfunctions in each
energy level are the Jost solutions $\psi_{k\pm}$. These are
determined by their asymptotic behavior that is given by
    \be
    \psi_{k\pm}(x)\to e^{\pm ikx}~~~~{\rm for}~~~~x\to\pm\infty.
    \label{jost}
    \ee
If we respectively denote the coefficients $A_\pm^\fg$ and
$B_\pm^\fg$ for the Jost solutions $\psi_{k\pm}$ as $A^\pm_\pm$ and
$B^\pm_\pm$, then in view of (\ref{asmp}) -- (\ref{jost}) we find
\cite{jpa-2009}
    \be
    A_+^+=B_-^-=1,~~~A_-^-=B_+^+=0,~~~A_-^+=B_+^-=M_{22},~~~
    A_+^-=M_{12},~~~B_-^+=-M_{21},
    \label{e1}
    \ee
where $M_{ij}$ are the entries of $\mathbf{M}$. Equations~\ref{e1}
show that the $\psi_{k\pm}$ are nothing but the left- and
right-going scattering solutions \cite{muga}, and that the left and
right transmission $T^{l,r}$ and reflection $R^{l,r}$ amplitudes are
given by \cite{p89}
    \be
    T^l=T^r=1/M_{22},~~~R^l=-M_{21}/M_{22},~~~R^r=M_{12}/M_{22}.
    \label{amplitudes}
    \ee

Spectral singularities of $H$ are eigenvalues $E_\star=k_\star^2$ at
which the Jost solutions become linearly-dependent
\cite{naimark,ss-math}. This happens if and only if $M_{22}=0$,
\cite{jpa-2009}. It is easy to see from (\ref{amplitudes}) that both
the transmission and reflections coefficients diverge at a spectral
singularity \cite{p89}. The latter condition is a characteristic
property of resonances, so a spectral singularity may be identified
with a peculiar type of a resonance that has a vanishing width. This
is because unlike ordinary resonances, the eigenvalue associated
with such a resonance is real. As mentioned earlier, the resonance
effect related with spectral singularity can be realized in certain
electromagnetic waveguides. These waveguides can be used to amplify
incoming waves with frequencies close to that of a spectral
singularity. Therefore they operate as resonators at these
frequencies \cite{p89}.

In the remainder of this section we examine the possibility of the
realization of the above resonance effect for the complex
delta-function potential:
    \be
    v(x)=z\,\delta(x),~~~~~z\in\C.
    \label{delta}
    \ee
The solution of the time-independent Schr\"odinger equation,
$H\psi=k^2\psi$, for this potential is elementary, \cite{jpa-2006b}.
We can use this solution, to determine the transfer matrix
$\mathbf{M}$. This yields
    \be
    \mathbf{M}=\left(\begin{array}{cc}
    1-\frac{iz}{2k} & -\frac{iz}{2k}\\
    \frac{iz}{2k} & 1+\frac{iz}{2k}\end{array}\right).
    \label{M=}
    \ee
Therefore $M_{22}=0$ if and only if $z=2ik$. This condition can be
satisfied for a real $k$ only if $z$ is imaginary and $k=-iz/2$.
Therefore, as noted in \cite{jpa-2006b} a spectral singularity
arises only for imaginary coupling constants. Furthermore, in view
of (\ref{amplitudes}) and (\ref{M=}), we have \footnote{The fact
that $T^l=T^r$ is generally true \cite{ahmed,p89}. The equality
$R^l=R^r$ arises from the fact that the delta-function potential is
even ($\cP$-symmetric).}
    \be
    T:=T^l=T^r=\frac{2k}{2k+iz},
    ~~~~~R:=R^l=R^r=\frac{-iz}{2k+iz}.
    \label{amplitudes2}
    \ee
In particular,
    \be
    |T|^2+|R|^2=
    \left(1-\frac{4k\Im(z)}{4k^2+|z|^2}\right)^{-1},
    \label{coeff}
    \ee
where $\Im$ stands for the imaginary part of its argument.

Clearly, for the case that $z$ is real the right-hand side of
(\ref{coeff}) is equal to unity. As is well-known, this is a
manifestation of the unitarity of the time-evolution with respect to
the $L^2$-inner product. For the cases that $z$ is not real,
$|T|^2+|R|^2$ deviates from unity. A spectral singularity
corresponds to the extreme situation where this quantity diverges.

Next, suppose that $z$ is imaginary, i.e., $z=i\lambda$ for some
$\lambda\in\R-\{0\}$, and $\epsilon:=1-\lambda/(2k)$ so that the
spectral singularity corresponds to $\epsilon=0$. Then
(\ref{amplitudes2}) and (\ref{coeff}) take the form
    \be
    T=R+1=\frac{1}{\epsilon},~~~~
    |T|^2+|R|^2=\frac{2(1-\epsilon)}{\epsilon^2}+1.
    \ee
In particular, the spectral singularity of this potential
corresponds to a quadratic divergence of $|T|^2+|R|^2$.

\section{Concluding Remarks}

The pioneering work of Bender and Boettcher on the reality of the
spectrum of $\cP\cT$-symmetric potentials such as $v(x)=ix^3$
generated a great deal of enthusiasm among theoretical physicists
who had mostly distanced themselves from non-Hermitian Hamiltonians
and complex potentials. This enthusiasm led to an extensive research
activity on the subject and produced a very large number of
publications. Most of these involve the study of various toy models
sharing the spectral properties of the imaginary cubic potential. It
soon became clear that the non-Hermitian Hamiltonians $H$ defined by
these potentials could not be used to model fundamental
(non-effective) physical processes unless one defined an inner
product that restored their Hermiticity. The existence of such inner
products and a basic method for their construction were obtained as
byproducts of a study of the mathematical structure behind the
appealing spectral properties of these operators \cite{p1,p2,p3}. In
fact, all that is needed is a metric operator $\etap$ that satisfies
the pseudo-Hermiticity relation $H^\dagger=\etap H\,\etap^{-1}$. All
the ingredients of the formalism are determined in terms of $\etap$
and independently of the choice of a $\cC$ operator.

For the cases that $H$ has an exact $\cP\cT$-symmetry, one can use
the prescription based on the $\cC$ operator, \cite{bender-review}.
This involves substituting the ansatz $\cC=e^Q\cC$ in $\cC^2=1$,
$[\cC,H]=0$, and $[\cC,\cP\cT]=0$, solving the resulting operator
equations for $Q$, and defining the physical Hilbert space using the
$\cC\cP\cT$-inner product that is identical with
$\br\cdot,\cdot\kt_{e^{-Q}}$.

This prescription has the disadvantage of relying on the
construction of a $\cC$ operator that does not enter the calculation
of the physical quantities. Furthermore, to employ it one needs to
deal with three operator equations. A more important drawback is
that this approach cannot be applied for systems that lack a
manifest antilinear symmetry. A typical example is the complex
delta-function potential $v(x)=z\delta(x)$. For the case that $z$
has a positive real part, the spectrum of
$-\frac{d^2}{dx^2}+z\delta(x)$ is purely real and continuous. It is
also free of spectral singularities. Hence one can apply the methods
of pseudo-Hermitian quantum mechanics \cite{review} to ``Hermitize''
the Hamiltonian $-\frac{d^2}{dx^2}+z\delta(x)$ and explore the
physical aspects of the quantum system it describes
\cite{jpa-2006b}. The approach based on the $\cC$ operator cannot be
applied to this system, because a priori the nature of the
underlying nonlinear symmetry of this system (the generalized
$\cP\cT$-symmetry \cite{jmp-2003}) is not known.

The study of quasi-Hermitian Hamiltonians that lack
$\cP\cT$-symmetry has been crucial in understanding the role and
meaning of spectral singularities. In this article, I reviewed the
essential features of spectral singularities and used the complex
delta-function potential to demonstrate how spectral singularities
appear as degeneracies of the reflection and transmission
coefficients. They are naturally interpreted as resonances with a
vanishing width. An experimental realization of the ensuing
resonance effect will be one of the rare instances of a physical
discovery that owes its existence to the recent study of
non-Hermitian Hamiltonian operators with a real spectrum.

\section*{Acknowledgments} This work has been supported by the Scientific and
Technological Research Council of Turkey (T\"UB\.{I}TAK) in the
framework of the project no: 108T009, and by the Turkish Academy of
Sciences (T\"UBA).

\end{document}